**Database Construction for Two-Dimensional Material-Substrate Interfaces**

Xian-Li Zhang(张现利)[1,2,#], Jinbo Pan(潘金波)[1#], Xin Jin(金鑫)[1,2], Yan-Fang Zhang(张艳芳)[2,1], Jia-Tao Sun(孙家涛)[3], Yu-Yang Zhang(张余洋)[2,4] , and Shixuan Du(杜世萱)[1,2,4,5*]

[1]Beijing National Laboratory for Condensed Matter Physics, and Institute of Physics, Chinese Academy of Sciences, Beijing 100190, China

[2]School of Physics, University of Chinese Academy of Sciences, Beijing 100049, China

[3]School of Information and Electronics, MIIT Key Laboratory for Low-Dimensional Quantum Structure and Devices, Beijing Institute of Technology, Beijing 100081, China

[4]CAS Center for Excellence in Topological Quantum Computation, University of Chinese Academy of Sciences, Beijing 100190, China

[5]Songshan Lake Materials Laboratory, Dongguan, Guangdong 523808, China

[#]Xian-Li Zhang and Jinbo Pan contributed equally to this work.

[*]Shixuan Du: sxdu@iphy.ac.cn

**Abstract**

The interfacial structures and interactions of two-dimensional (2D) materials on solid substrates are of fundamental importance for the fabrication and application of 2D materials. However, selection of a suitable solid substrate to grow 2D material, determination and control of the 2D material-substrate interface remain a big challenge due to the large diversity of possible configurations. Here, we propose a computational framework to select an appropriate substrate for epitaxial growth of 2D material and to predict possible 2D material-substrate interface structures and orientations using density functional theory calculations performed for all non-equivalent atomic structures satisfying the symmetry constraints. The approach was validated by the correct prediction of three experimentally reported 2D material-substrate interface systems with only the given information of two parent materials. Several possible interface configurations are also proposed based on this approach. We therefore construct a database that contains these interface systems and has been continuously expanding. This database serves as preliminary guidance for epitaxial growth and stabilization of new materials in experiments.

In past decades, various high-throughput calculation frameworks have been designed to discover materials with superior and tailored properties.[1-15] They are now becoming increasingly important for the discovery of new functional materials. The automated FLOW (AFLOW)[16, 17] package provides a groundbreaking software framework for high-throughput calculations of structural and electronic properties of inorganic crystals. The Materials Project[18, 19] is another milestone materials database containing hundreds of thousands of compounds. Due to distinct mechanical and electronic properties, 2D materials have attracted broad interest, and many high-throughput calculations also have been performed.[20-27] More than a thousand 2D materials, which are easily or potentially exfoliatable from bulk materials, were reported by Mounet.[24] The Computational 2D materials Database (C2DB),[27] which is developed by Hasstrup, contains the stability, electronic, magnetic and optical properties of thousands of 2D materials distributed across tens of different crystal structures. All these databases provide convenient tools to search for novel functional materials in batteries, electrocatalysts, etc.

Substrates, most of the time, play non-negligible roles in the fabrications and applications of 2D materials. For a 2D material epitaxially grown on a substrate, the lattice mismatch is typically unavoidable, leading to the lattice dilation and compression of the component.[28] Meanwhile, the physical properties of these atomically thin materials are usually sensitive to strain.[29-32] In addition, the coupling between a 2D material (adsorbate) and a substrate may affect the intrinsic physical properties of the 2D material. For example, the Dirac cone of graphene (Gr) is distorted while grown on some metallic or semiconducting substrates.[33-35] However, in previous work, the statistical effects from the supporting substrates including atomic configurations and electronic properties have not yet been fully considered,[28, 33-38] although all the effects coming from the specific configurations of adsorbates on a substrate (e.g., twist angle, strain, buckled height) are ubiquitous, which hinder the further exploration of new material on the substrate.

We noticed that there are several group reported high-throughput calculations of interfacial systems including solid/solid hetero-structures, solid/implicit solvents systems, nanoparticle/ligands systems,[39] computational framework to select optimal substrates for epitaxial growth of polymorphs,[40] and a method for predicting solid/solid interface structures.[41] However, the database of 2D materials on the substrates is still in its infancy.

In this paper, we present a high-throughput calculation workflow aiming to predict the configurations of 2D materials on substrates beyond experiments. A series of unique and irreducible interfacial structures were generated under biaxial strains and symmetry constraints. High-throughput density functional theory (DFT) calculations were performed to obtain the optimized structures and binding energies. The stable configurations that can be synthesized experimentally are predicted based on their binding energies. This method has been proven valid by various 2D-material-substrate interfaces, including Gr@Ru(0001),[42-44] Gr@Ir(111),[45, 46] arsenene@Ag(111),[47] and extended to 2D heterostructure antimonene@PdTe$_2$,[48] which have been reported in experiments. Based on these successful examples, a database,[49] including several new materials, is created. The calculations on new configurations containing 2D material/substrate interfaces and 2D-2D interfaces are ongoing and will be included in the database in the future. We believe these results will improve our ability to explore and design 2D material on substrates and can be extended to construct 2D heterostructures which exhibit new and unique properties.

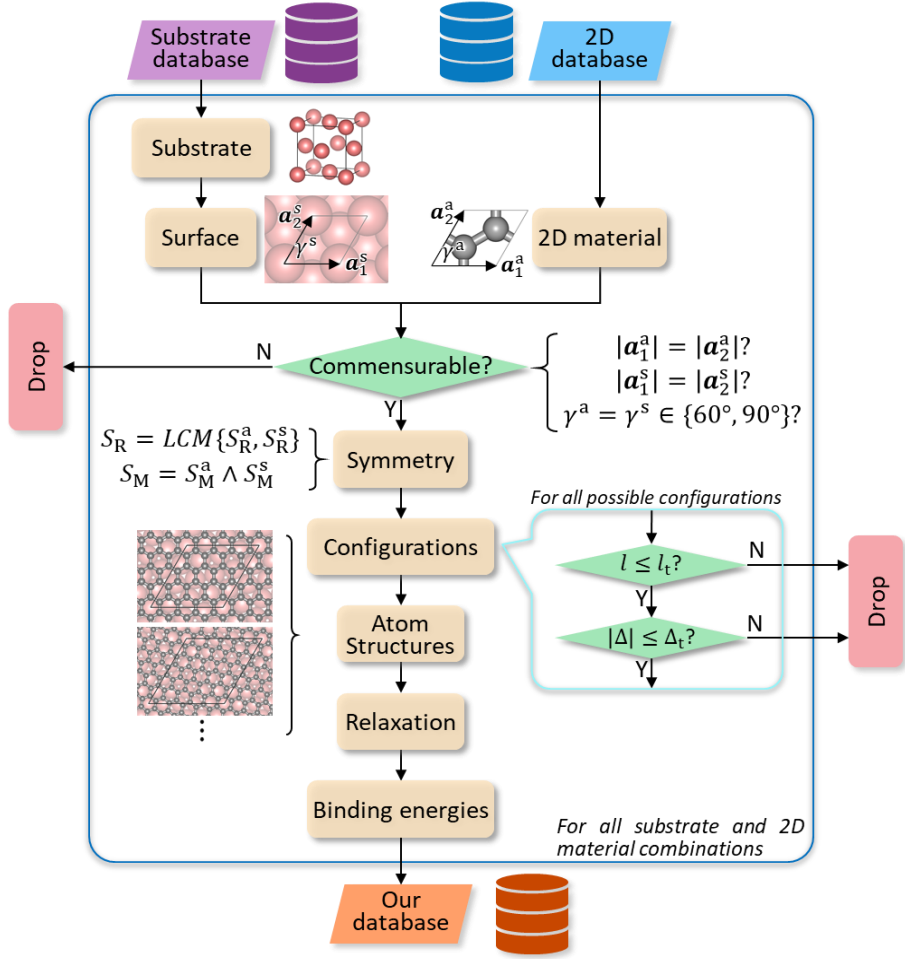

**Figure 1.** Workflow for 2D-surface interface structures prediction. The interfaces of the substrate and the adsorbate, selected from the two input databases, were filtered by the commensurability and symmetry requirements. The DFT calculations were performed to obtain the stable structures and binding energies.

The workflow for the interface database construction is illustrated in Figure 1. Two input databases, which contain the potential 2D materials, and substrates were constructed from an existing materials database, i.e., C2DB,[27] Materials Project.[18, 19]

At the start of the workflow, the adsorbate (a 2D material) and a substrate were chosen from the input databases. The bulk phase of a substrate was cleaved along a high Miller index surface to fit the adsorbate. In fact, 2D material tends to form more uniform sheets on a symmetry matched surface.[50-52] The 2D material is compressed or stretched biaxially to fit the lattice of the substrate and multiple twist angles are also allowed. Under these prerequisites, the lattice of the cleaved substrate and adsorbate should meet

(shown in Figure 1)

$$\begin{cases} |a_1^a| = |a_2^a| \\ |a_1^s| = |a_2^s| \\ \gamma_a = \gamma_s \in \{60°, 90°\} \end{cases}$$

where $|a_1^a|$ and $|a_2^a|$ are the vectors length of the adsorbate, $|a_1^s|$ and $|a_2^s|$ are the vectors length of the substrate, $\gamma^a$ and $\gamma^s$ are the angles between the two lattice vectors of the adsorbate and the substrate, respectively. That is, the substrate and the adsorbate should be the same 2D Bravais lattices, hexagonal or square. In the following, we use the hexagonal lattices as examples.

We then generated the superlattices for the adsorbate and substrate, respectively. Due to the constraints of the commensurate supercell, the index notation $\boldsymbol{M} = (m, n)$ is used to describe the superlattice, where $m$ and $n$ are two integers. The details about the index notation are discussed in the supplementary material. The relative length $L$ can be expressed by $L = |\bar{a}_1|/|a_1|$, where $|\bar{a}_1|$ is the vector length of supercell. The large value of irrational $L$ is rarely reported in previous literature. Thus, we only kept those configurations with integer $L$ when $L$ is greater than $\sqrt{50}$ in the following.

We then considered the constraint condition of the symmetries of the adsorbate and the substrate. Different from a 3D lattice, only the rotation axes perpendicular to the surface and the mirror plane normal to the surface are allowed in the 2D system. If the mirror symmetry does not exist in any of the two materials, the effective range of twist angle is described as $\left[0°, \frac{360°}{LCM(S_R^a, S_R^s)}\right)$, where $LCM(S_R^a, S_R^s)$ is the lowest common multiple (LCM) of rotation symmetries of the adsorbate ($S_R^a$) and the substrate ($S_R^s$) (shown in Table S1 and Figure S2). In contrast, if the mirror symmetry both exists in the substrate and the adsorbate, the effective range of the twist angle could further reduce to $\left[0°, \frac{180°}{LCM(S_R^a, S_R^s)}\right]$. It should be noted that the range of the twist angle is a right-open interval in the former case and a close interval in the latter.

Subsequently, a series of configurations are generated by combining all possible superlattices of the adsorbate and the substrate. However, the simple combination operation takes a large number of configurations. Some of them are unreasonable or computationally expensive. To balance the computational accuracy and cost, the oversized system was excluded when the lattice constant $l$ is larger than the limitation $l_t$ (typical value is 30 Å). It should be noted that we assume that the rigid substrate is not strained, i.e., the vectors lengths of the commensurate structure are equal to those of the superlattice of the substrate ($l \equiv |\bar{\boldsymbol{a}}_1^s|$).

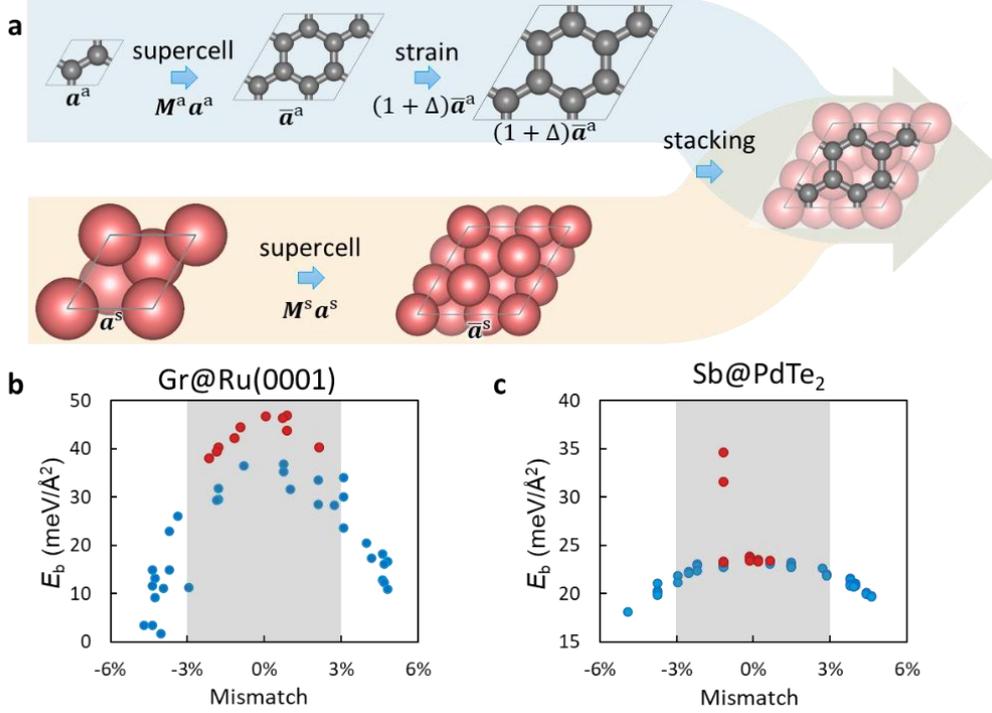

**Figure 2.** The construction of 2D material and substrate interface. (**a**) An illustration of building supercells under $\Delta$ strains and stacking processes. The binding energy $E_b$ vs lattice mismatch for the configurations of (**b**) Gr@Ru(0001) and (**c**) Sb@PdTe$_2$. The red spots represent the ten most stable configurations.

We also screened the configurations with a large lattice mismatch. In the definition of mismatch $\Delta = (l - |\bar{\boldsymbol{a}}_1^a|)/|\bar{\boldsymbol{a}}_1^a| = (|\bar{\boldsymbol{a}}_1^s| - |\bar{\boldsymbol{a}}_1^a|)/|\bar{\boldsymbol{a}}_1^a|$, $\Delta$ is equal to the strain of the adsorbate (Figure 2a). That is, a negative value of $\Delta$ implies the adsorbate is compressed and vice versa. In general, materials will be unstable under large strains. We plotted the $E_b$ vs $\Delta$ of several configurations in Figure 2b-c. The normalized binding energy is defined as $E_b = (E^a + E^s - E^c)/S$, where the $S$ is the area of the superlattice, and the $E^a$, $E^s$ and $E^c$ are energies of the adsorbate, the substrate and their composite structure, respectively. The configurations having larger binding energies are more stable according to the definition. There is an approximate negative correlation between

$E_b$ and the absolute value of mismatch $|\Delta|$. The $|\Delta|$ of the most stable 10 configurations (red points) are always less than 3% (gray area region). Therefore, we drop the likely unstable configurations if $|\Delta| < \Delta_t$, where the $\Delta_t$ is the limitation of mismatch. Interestingly, the most stable configuration could deviate from the smallest mismatch (shown in Figure 2b-c). Thus, the smallest $\Delta$ is insufficient to guarantee the most stable configuration.

The stacking process is illustrated in Figure 2a. The superlattices of an adsorbate or a substrate are built by multiplying their primitive cell with a superlattice matrix. Furthermore, the superlattice of the adsorbate is multiplied by $(1+\Delta)$ to match the magnitude of the substrate before stacking the superlattices of the adsorbate and the substrate to obtain a commensurate structure. Moreover, different adsorption sites were considered on the substrate (shown in Figure S3a). In fact, several adsorption sites are implicitly included in the large supercell (shown in Figure S3b). Thus, we only construct different structures to include all the possible adsorption sites for the small superlattices (typically for $l < 15$ Å).

Finally, we performed structural optimizations and total energy calculations for each atomic structure based on DFT calculations. The optimized structure and binding energy of configurations were added to the database. The configurations were sorted by the binding energy normalized to the area. For the interface systems that have several configurations with significantly larger binding energies than others, we consider that they are the stable configurations. Otherwise, the configurations are deemed to be stable if the normalized binding energy is larger than 80% of the most stable structure.

**Table 1.** Top 10 configurations with large binding energies, $E_b$, of Gr@Ru(0001)[a]

| $M^{Gr}$ | $M^{Ru(0001)}$ | Gr@Ru(0001) | $l$ (Å) | $\Delta$ | $\theta$ (°) | $E_b$ (meV/Å2) |
|---|---|---|---|---|---|---|
| (12, 0) | (11, 0) | 12@11 | 29.87 | 0.89% | 0.00 | 47 |
| (11, 0) | (10, 0) | 11@10 | 27.15 | 0.06% | 0.00 | 47 |
| (13, 0) | (12, 0) | 13@12 | 32.58 | 1.60 % | 0.00 | 46 |
| (6, 1) | (6, 0) | √43@6 | 16.29 | 0.71% | 7.59 | 46 |
| (10, 0) | (9, 0) | 10@9 | 24.44 | -0.94% | 0.00 | 44 |
| (5, 0) | (1, 4) | 5@√21 | 12.44 | 0.88% | 10.89 | 44 |
| (5, 1) | (5, 0) | √31@5 | 13.58 | -1.16% | 8.95 | 42 |
| (6, 0) | (1, 5) | 6@√31 | 15.12 | 2.14% | 8.95 | 40 |
| (5, 3) | (5, 2) | 7@√39 | 16.96 | -1.81% | 5.68 | 40 |
| (5, 2) | (5, 1) | √39@√31 | 15.12 | -1.87% | 7.15 | 39 |
| (9, 0) | (8, 0) | 9@8 | 21.72 | -2.16% | 0.00 | 38 |

[a] $M^{Gr}$ and $M^{Ru(0001)}$ are the superlattice matrices of the adsorbate and substrate, respectively. The experimentally reported configurations are labeled in orange background. The $l$ is the lattice constant of supercell, and $\Delta$ and $\theta$ are the mismatch and the twist angle between the adsorbate and substrate, respectively.

We demonstrate our workflow in graphene on Ru(0001), which is an extensively investigated system.[42, 44, 53-57] The first ten configurations with largest binding energies were presented in Table 1. The first three of them, i.e., (12, 0)@(11, 0), (11, 0)@(10, 0) and (13, 0)@(12, 0), have been fabricated experimentally in previous work.[44, 53, 54] Here, the indexes before and after "@" correspond to the superlattice index of adsorbate and substrate, respectively. The configuration of a 25 × 25 graphene honeycomb sitting on a 23 × 23 ruthenium far exceeds the threshold of lattice constant $l_t$, thus has not been considered in our calculations. However, the configuration (25, 0)@(23, 0) could be regarded as the combination of (13, 0)@(12, 0) and (12, 0)@(11, 0). Therefore, the configuration of (13, 0)@(12, 0) with the lattice constant (32.6 Å) is slightly larger than $l_t$, was calculated as an exception. These agreements make it reasonable to use the binding energy as the criterion for predicting the experimentally possible configurations of 2D materials on substrates. Interestingly, we found some configurations (e.g., (6, 1)@(6, 0), (5, 0)@(1, 4), *etc*.) with large binding energies and non-zero twist angle that have not been reported so far. We believe some of these configurations also have great potential to be observed if more experiments have been done. For some other systems, nucleation centers have preferred orientations, which may also affect the final configurations. Nevertheless, the predicted configuration with relatively strong binding energy may be obtained by a specific method, e.g., high-temperature annealing, etc.

**Table 2.** Top 20 configurations with large binding energies of Gr@Ir(111)[a]

| $M^{Gr}$ | $M^{Ir(111)}$ | Gr@Ir(111) | $l$ (Å) | Δ | $\theta$ (°) | $E_b$ (meV/Å$^2$) |
|---|---|---|---|---|---|---|
| (4, 0) | (1, 3) | 4@√13 | 9.87 | 0.05% | 13.90 | 38 |
| (5, 1) | (5, 0) | √31@5 | 13.69 | -0.33% | 8.95 | 37 |
| (10, 0) | (9, 0) | 10@9 | 24.64 | -0.11% | 0.00 | 37 |
| (5, 3) | (5, 2) | 7@√39 | 17.10 | -0.98% | 5.68 | 36 |
| (7, 0) | (2, 5) | 7@√39 | 17.10 | -0.98% | 16.10 | 35 |
| (5, 2) | (5, 1) | √39@√31 | 15.25 | -1.05% | 7.15 | 35 |
| (3, 5) | (5, 2) | 7@√39 | 17.10 | -0.98% | 22.11 | 35 |
| (5, 2) | (1, 5) | √39@√31 | 15.25 | -1.05% | 25.05 | 35 |
| (11, 0) | (10, 0) | 11@10 | 27.38 | 0.90% | 0.00 | 34 |
| (9, 0) | (8, 0) | 9@8 | 21.91 | -1.34% | 0.00 | 34 |
| (6, 1) | (6, 0) | √43@6 | 16.43 | 1.56% | 7.59 | 31 |
| (3, 4) | (5, 1) | √37@√31 | 15.25 | 1.59% | 25.77 | 30 |
| (4, 3) | (5, 1) | √37@√31 | 15.25 | 1.59% | 16.34 | 30 |
| (5, 0) | (1, 4) | 5@√21 | 12.55 | 1.72% | 10.89 | 29 |
| (3, 2) | (4, 0) | √19@4 | 10.95 | 1.85% | 23.41 | 29 |
| (3, 0) | (1, 2) | 3@√7 | 7.24 | -2.12% | 19.11 | 25 |
| (4, 4) | (4, 3) | √48@√37 | 16.66 | -2.55% | 4.72 | 23 |
| (8, 0) | (7, 0) | 8@7 | 19.17 | -2.88% | 0.00 | 19 |
| (1, 6) | (4, 3) | √43@√37 | 16.66 | 2.96% | 27.13 | 16 |
| (1, 6) | (3, 4) | √43@√37 | 16.66 | 2.96% | 17.70 | 16 |

[a]The experimentally reported configurations are labeled in the orange background.

We also demonstrate the workflow in graphene on a Ir(111) surface, which is a medium interacting system. Due to the relative weak van der Waals (vdW) bonding, graphene shows multi-oriented superstructures on the Ir(111) surface.[45, 46] The two most stable configurations predicted by high-throughput calculations in this work have been observed experimentally (shown in Table 2). The energy ranking $R$ of rest experimental results are in the range from 11 to 20 (shown in Table S2). It is worth noting that the (2, 0)@(1,1) configuration possesses a negative binding energy, implying the energetically unfavorable, which is in contrast to experimental results.[45, 46] Moreover, there is also a series of large binding energy configurations, which have the potential to be fabricated in the future.

It is well known that the choice of exchange-correlation functional is important for systems with medium and weak interaction, i.e., the adsorption height and binding energy will be significantly affected. Nevertheless, stable configurations predicted by the high-throughput calculation method are independent of the choice of exchange-correlation function. The robustness of our method was checked by testing different exchange-correlation functionals. The binding energy and their ranking $R$ of three different functionals (i.e., DFT-D3, LDA and optB88-vdW) have been listed in Table S2. The binding energy of the specific configuration obtained using different functionals is significantly different, but their ranking shows negligible dependence on the functionals used.

**Table 3.** Top 10 configurations with large binding energies of Sb@PdTe$_2$[a]

| $M^{Sb}$ | $M^{PdTe_2}$ | Sb@PdTe$_2$ | $l$ (Å) | Δ | $\theta$ (°) | $E_b$ (meV/Å$^2$) |
|---|---|---|---|---|---|---|
| (1, 0) | (1, 0) | 1@1 | 4.07 | -1.17% | 0.00 | 35 |
| (0, 1) | (1, 0) | 1@1 | 4.07 | -1.17% | 60.00 | 32 |
| (-4, 8) | (3, 5) | √48@7 | 28.49 | -0.15% | 51.79 | 24 |
| (4, 4) | (7, 0) | √48@7 | 28.49 | -0.15% | 30.00 | 24 |
| (0, 6) | (4, 3) | 6@√37 | 24.75 | 0.19% | 34.72 | 24 |
| (-3, 6) | (2, 4) | √27@√28 | 21.53 | 0.64% | 49.11 | 23 |
| (-4, 4) | (5, 3) | √48@7 | 28.49 | -0.15% | 8.21 | 23 |
| (-3, 7) | (3, 4) | √37@√37 | 24.75 | -1.17% | 50.57 | 23 |
| (0, 6) | (3, 4) | 6@√37 | 24.75 | 0.19% | 25.28 | 23 |
| (-3, 8) | (3, 5) | 7@7 | 28.49 | -1.17% | 43.57 | 23 |

[a]The experimentally reported configurations are shown with an orange background and they have significantly higher binding energy than other configurations.

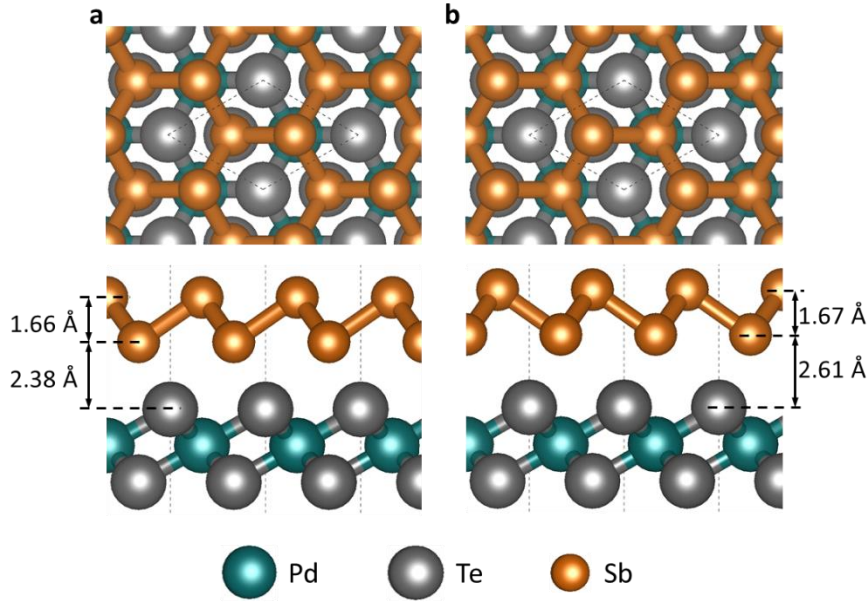

**Figure 3.** The two most stable structures of Sb@PdTe$_2$ predicted by high-throughput calculations. The configuration of **(a)** $(0,1)@(1,0) - R60°$ was reported in the previous work[48] and the configuration of **(b)** $(1,0)@(1,0) - R0°$ is another possible configuration predicted by our high-throughput calculations.

Our workflow is also applicable to lower-symmetry systems. Unlike graphene, freestanding group-VA monolayers have a buckled honeycomb structure, which holds the three-fold rotational symmetry. The available twist angle $\theta$ is in the range of [0°, 60°] (shown in Table S1). We calculated the possible configurations of buckled antimonene (b-Sb) on a PdTe$_2$ substrate and presented the results in Table 3. The PdTe$_2$ is one of the typical layered materials. Therefore, we considered the (001) surface as the substrate, which is not specially labeled in the following. The two largest superstructures with the largest binding-energy, which are $(1, 0)@(1,0)$ and $(0, 1)@(1, 0)$ (shown in Figure 3), have significant larger binding energies (35 and 32 meV/Å$^2$) than other configurations ($\leq$ 24 meV/Å$^2$). Therefore, both structures are grouped into the most stable configurations. It is noted that these two composite structures cannot be transformed into each other through symmetry operation or changing the adsorption site. That is, two different orientations of antimonene with the same period could form on the PdTe$_2$ surface. However, only $(0, 1)@(1,0)$ was reported in the previous work.[48] As $(1, 0)@(1,0)$ and $(0, 1)@(1, 0)$ yield the same scanning tunneling microscope (STM) images and low energy electron diffraction (LEED) pattern, we believe that the (1, 0)@(1, 0) may also exist but may not have been identified correctly in Ref [44]. It is worthy of noting that the 2D materials heterostructures could be predicted by considering 2D phase $PdTe_2$. In this case, all atoms of 2D $PdTe_2$ (contrastively, only one atom layer in bulk $PdTe_2$) was relaxed.

The same situation exists in the arsenene on Ag(111) system. The 4×4 buckled arsenene on 5×5 Ag(111) was fabricated experimentally.[47] A series of linear boundaries were interpreted as the small shift of the domains. However, considering the three-fold rotational symmetry in both of the buckled arsenene (b-As) and the Ag(111) surface, the two nonequivalent configurations of (4, 0)@(5, 0) and (0, 4)@(5, 0) (as shown in Table S3), with similar binding energies, can be expected to co-exist. The experimentally observed linear bright features are more likely to be the grain boundary of these two configurations.

We also evaluated the possibility of flattened arsenene epitaxially grown on the substrate, which is an analogue to planar graphene. We compared the buckled arsenene (b-As) and the flat arsenene (f-As) on an Ag(111) substrate (shown in Table S3 and Table S4). It is found that the binding energies of the buckled arsenene are much larger than those of the flat arsenene indicating that the b-As is energetically favorable to be synthesized on the Ag(111) substrate.

It should be noted that tens of configurations of b-As@Ag(111) show similar binding energies (shown in Table S3). Among them, only the (4, 0)@(5, 0) and (0, 4)@(5, 0) of b-As@Ag(111) were synthesized at specific experimental conditions. Other possible configurations (shown in Table S3) also have the potential to be fabricated on the Ag(111) substrate. On the other hand, however, several different configurations which are energetically-favorable, signifying multiple phases of arsenene, could also be formed on the Ag(111) surface, and more precise experimental conditions are required to fabricate the specific large-scale monocrystalline structure.

We built a computational database predicting the atomic structures of the 2D material-substrate interface. High-throughput DFT calculations were performed to get the energetically favorable configurations. The validity of this method is verified by several reported systems, e.g., Gr@Ru(0001), Gr@Ir(111), Sb@$PdTe_2$, etc. Based on this method, we predicted several possible configurations which could be potentially synthesized in experiments. The database provides a convenient and powerful tool for the discovery of suitable substrates to grow 2D materials and novel 2D heterostructures.

**Footnotes**

We acknowledge financial support from the National Key R&D program of China (Nos. 2019YFA0308500, 2020YFA0308800 and 2016YFA0202300), National Natural Science Foundation of China (Nos. 51922011, 61888102, and 11974045), Strategic Priority Research Program of the Chinese Academy of Sciences (Nos. XDB30000000 and XDB28000000), Beijing Institute of Technology Research Fund Program for Young Scholars, and the Fundamental Research Funds for the Central Universities.

**Table 1.** Top 10 configurations with large binding energies, $E_b$, of Gr@Ru(0001)[a]

| $M^{Gr}$ | $M^{Ru(0001)}$ | Gr@Ru(0001) | $l$ (Å) | Δ | $\theta$ (°) | $E_b$ (meV/Å²) |
|---|---|---|---|---|---|---|
| (12, 0) | (11, 0) | 12@11 | 29.87 | 0.89% | 0.00 | 47 |
| (11, 0) | (10, 0) | 11@10 | 27.15 | 0.06% | 0.00 | 47 |
| (13, 0) | (12, 0) | 13@12 | 32.58 | 1.60 % | 0.00 | 46 |
| (6, 1) | (6, 0) | √43@6 | 16.29 | 0.71% | 7.59 | 46 |
| (10, 0) | (9, 0) | 10@9 | 24.44 | -0.94% | 0.00 | 44 |
| (5, 0) | (1, 4) | 5@√21 | 12.44 | 0.88% | 10.89 | 44 |
| (5, 1) | (5, 0) | √31@5 | 13.58 | -1.16% | 8.95 | 42 |
| (6, 0) | (1, 5) | 6@√31 | 15.12 | 2.14% | 8.95 | 40 |
| (5, 3) | (5, 2) | 7@√39 | 16.96 | -1.81% | 5.68 | 40 |
| (5, 2) | (5, 1) | √39@√31 | 15.12 | -1.87% | 7.15 | 39 |
| (9, 0) | (8, 0) | 9@8 | 21.72 | -2.16% | 0.00 | 38 |

[a] $M^{Gr}$ and $M^{Ru(0001)}$ are superlattice matrices of the adsorbate and substrate, respectively. The experimentally reported configurations are labeled in the orange background. The $l$ is the lattice constant of the supercell, and Δ and $\theta$ are the mismatch and the twist angle between the adsorbate and substrate, respectively.

**Table 2.** Top 20 configurations with large binding energies of Gr@Ir(111)[a]

| $M^{Gr}$ | $M^{Ir(111)}$ | Gr@Ir(111) | $l$ (Å) | Δ | $\theta$ (°) | $E_b$ (meV/Å²) |
|---|---|---|---|---|---|---|
| (4, 0) | (1, 3) | 4@√13 | 9.87 | 0.05% | 13.90 | 38 |
| (5, 1) | (5, 0) | √31@5 | 13.69 | -0.33% | 8.95 | 37 |
| (10, 0) | (9, 0) | 10@9 | 24.64 | -0.11% | 0.00 | 37 |
| (5, 3) | (5, 2) | 7@√39 | 17.10 | -0.98% | 5.68 | 36 |
| (7, 0) | (2, 5) | 7@√39 | 17.10 | -0.98% | 16.10 | 35 |
| (5, 2) | (5, 1) | √39@√31 | 15.25 | -1.05% | 7.15 | 35 |
| (3, 5) | (5, 2) | 7@√39 | 17.10 | -0.98% | 22.11 | 35 |
| (5, 2) | (1, 5) | √39@√31 | 15.25 | -1.05% | 25.05 | 35 |
| (11, 0) | (10, 0) | 11@10 | 27.38 | 0.90% | 0.00 | 34 |
| (9, 0) | (8, 0) | 9@8 | 21.91 | -1.34% | 0.00 | 34 |
| (6, 1) | (6, 0) | √43@6 | 16.43 | 1.56% | 7.59 | 31 |
| (3, 4) | (5, 1) | √37@√31 | 15.25 | 1.59% | 25.77 | 30 |
| (4, 3) | (5, 1) | √37@√31 | 15.25 | 1.59% | 16.34 | 30 |
| (5, 0) | (1, 4) | 5@√21 | 12.55 | 1.72% | 10.89 | 29 |
| (3, 2) | (4, 0) | √19@4 | 10.95 | 1.85% | 23.41 | 29 |
| (3, 0) | (1, 2) | 3@√7 | 7.24 | -2.12% | 19.11 | 25 |
| (4, 4) | (4, 3) | √48@√37 | 16.66 | -2.55% | 4.72 | 23 |
| (8, 0) | (7, 0) | 8@7 | 19.17 | -2.88% | 0.00 | 19 |
| (1, 6) | (4, 3) | √43@√37 | 16.66 | 2.96% | 27.13 | 16 |
| (1, 6) | (3, 4) | √43@√37 | 16.66 | 2.96% | 17.70 | 16 |

[a]The experimentally reported configurations are labeled in the orange background.

**Table 3.** Top 10 configurations with large binding energies of Sb@PdTe$_2$[a]

| $M^{Sb}$ | $M^{PdTe_2}$ | Sb@PdTe$_2$ | $l$ (Å) | Δ | $\theta$ (°) | $E_b$ (meV/Å$^2$) |
|---|---|---|---|---|---|---|
| (1, 0) | (1, 0) | 1@1 | 4.07 | -1.17% | 0.00 | 35 |
| (0, 1) | (1, 0) | 1@1 | 4.07 | -1.17% | 60.00 | 32 |
| (-4, 8) | (3, 5) | √48@7 | 28.49 | -0.15% | 51.79 | 24 |
| (4, 4) | (7, 0) | √48@7 | 28.49 | -0.15% | 30.00 | 24 |
| (0, 6) | (4, 3) | 6@√37 | 24.75 | 0.19% | 34.72 | 24 |
| (-3, 6) | (2, 4) | √27@√28 | 21.53 | 0.64% | 49.11 | 23 |
| (-4, 4) | (5, 3) | √48@7 | 28.49 | -0.15% | 8.21 | 23 |
| (-3, 7) | (3, 4) | √37@√37 | 24.75 | -1.17% | 50.57 | 23 |
| (0, 6) | (3, 4) | 6@√37 | 24.75 | 0.19% | 25.28 | 23 |
| (-3, 8) | (3, 5) | 7@7 | 28.49 | -1.17% | 43.57 | 23 |

[a]The experimentally reported configurations are shown with an orange background, they have significantly higher binding energy than other configurations.

**Figure 1.** Workflow for 2D-surface interface structures prediction. The interfaces of the substrate and the adsorbate, selected from the two input databases, were filtered by the commensurability and symmetry requirements. The DFT calculations were performed to obtain the stable structures and binding energies.

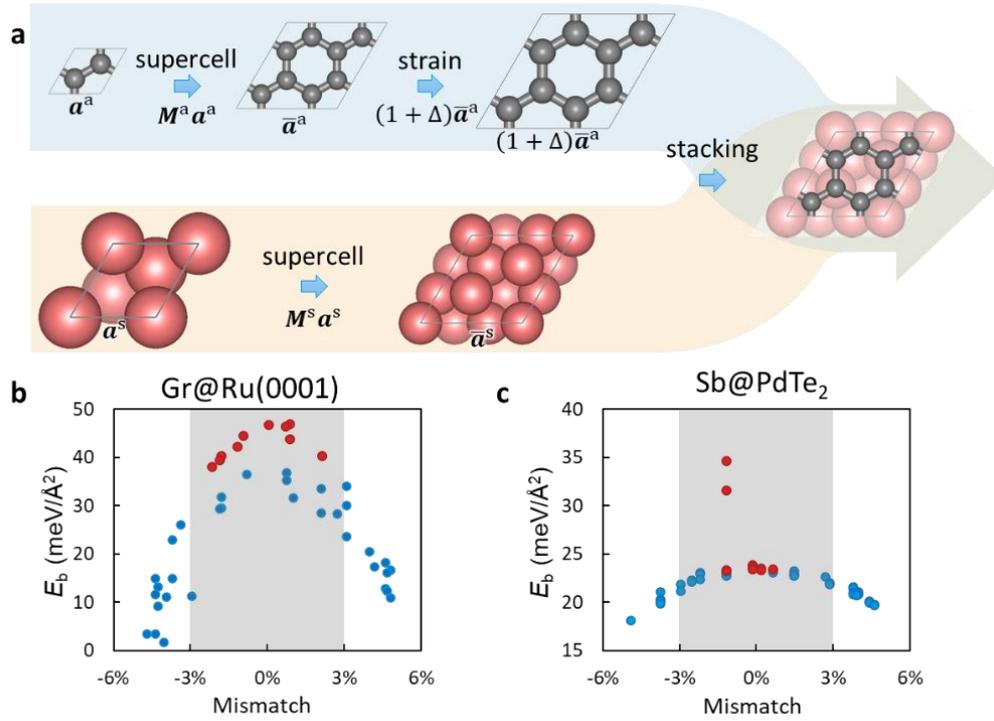

**Figure 2.** The construction of 2D material and substrate interface. (**a**) An illustration of building supercells under Δ strains and stacking processes. The binding energy $E_b$ vs lattice mismatch for the configurations of (**b**) Gr@Ru(0001) and (**c**) Sb@PdTe$_2$. The red spots represent the ten most stable configurations.

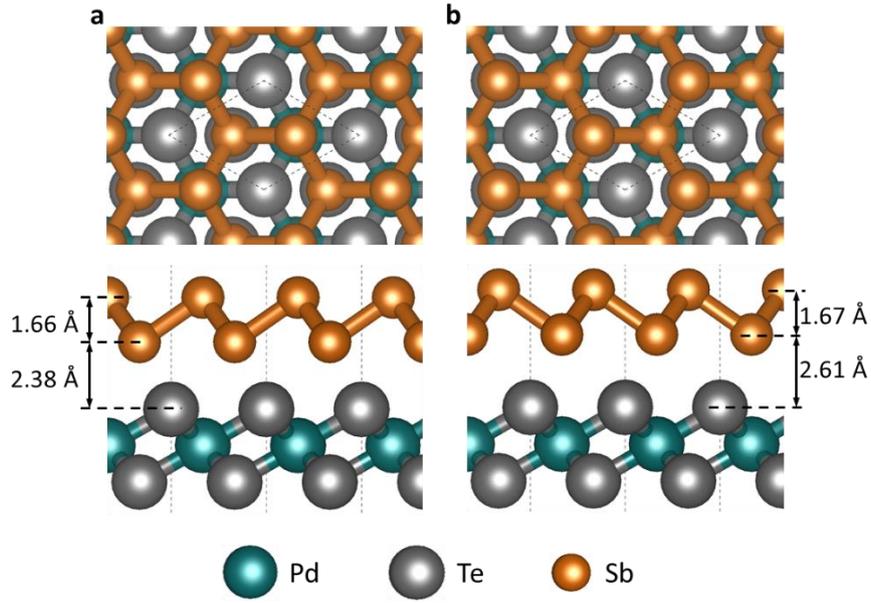

**Figure 3.** The two most stable structures of Sb@PdTe$_2$ predicted by high-throughput calculations. The configuration of **(a)** $(0, 1)@(1, 0) - R60°$ was reported in the previous work[48] and the configuration of **(b)** $(1, 0)@(1, 0) - R0°$ is another possible configuration predicted by our high-throughput calculations.